\documentclass[%
 reprint,
 amsmath,amssymb,
 aps,
floatfix,
nofootinbib]{revtex4-1}

\usepackage{graphicx}% Include figure files
\usepackage{dcolumn}% Align table columns on decimal pointa
\usepackage{bm}% bold math
\usepackage[symbol]{footmisc}

\usepackage{amsmath}
\usepackage{xcolor}
\usepackage{float}
\usepackage{multirow}
\usepackage{chemformula}

\usepackage{natbib}
\begin{document}

\preprint{APS/123-QED}

\title{Vacancy Diffusion Across FeCrAl Alloy Composition Space for Accident-Tolerant Fuel Cladding}

\author{Mihai Pitigoi}
\affiliation{Department of Physics and Astronomy, The University of Manchester, Oxford Road, Manchester M13 9PL, United Kingdom}

\author{Peter Hatton}
\affiliation{Amentum Clean Energy, 305 Bridgewater Place, Birchwood Park, Warrington, UK, WA3 6XF}

\date{\today}

\begin{abstract}

Iron-chromium-aluminium (FeCrAl) alloys are leading candidates for accident-tolerant fuel cladding in light-water reactors, where their superior high-temperature oxidation resistance promises to extend coping times during loss-of-coolant accidents. The in-reactor lifetime of cladding is ultimately governed by radiation-induced microstructural evolution of which point defect transport is the dominant mechanism, however, this remains poorly understood. Here, we use a species-resolved kinetic Monte Carlo (KMC) model for vacancy diffusion in FeCrAl, parameterised by linear surrogate models trained on a database of migration barriers generated through the Hop-Decorate workflow. By sampling compositions spanning the Fe-rich to Cr-rich range of the Fe-Cr-Al system, we map how the local chemical environment controls vacancy hopping and hence macroscopic diffusivity. We find that increasing the Cr content in the alloy progressively decreases global diffusivity of vacancies even though activation energies stay relatively constant. This implies that the higher the Fe content in the alloy, the faster vacancies diffuse, thereby increasing annihilation events with fast-moving interstitials, potentially reducing irradiation induced defects and increasing radiation tolerance. Conversely, we find that Cr-rich alloy compositions stand out with a markedly elevated activation energy and significantly slower diffusion, orders of magnitude lower at accident-relevant temperatures. This indicates suppressed vacancy mobility in Cr-rich $\alpha'$ phases which are known to form under irradiation.
\end{abstract}

\maketitle
\section{Introduction}

Accident-tolerant fuel (ATF) development has become a central priority in nuclear materials research following the Fukushima Daiichi accident, which exposed the vulnerability of zirconium-based claddings under extended loss-of-cooling conditions \cite{IAEA2015Fukushima}. The rapid oxidation of Zircaloy in high-temperature steam environments, producing hydrogen and compromising containment, demonstrated that next-generation cladding materials must provide improved oxidation resistance, enhanced mechanical stability, and predictable behaviour under extreme thermal and radiative loading.

The 2011 Fukushima Daiichi accident resulted from a station blackout following the T\=ohoku earthquake and tsunami, which disabled cooling systems across multiple reactor units \cite{IAEA2015Fukushima}. As decay heat could no longer be removed, fuel temperatures rose sharply, driving steam oxidation of the Zircaloy cladding via the exothermic reaction $\mathrm{Zr} + 2\mathrm{H_2O} \rightarrow \mathrm{ZrO_2} + 2\mathrm{H_2}$. This reaction is self-accelerating above approximately 1200$^\circ$C, and the hydrogen generated accumulated and ultimately ignited, causing building explosions that compromised containment and dispersed radioactive material \cite{IAEA2015Fukushima}. Critically, the failure was not caused by the reactor fuel itself but by the cladding's thermochemical instability under prolonged station blackout conditions, representing a scenario that accident-tolerant fuel programmes now explicitly target.

Iron-chromium-aluminium (FeCrAl) alloys have emerged as one of the most promising ATF cladding candidates, primarily due to their exceptional high-temperature oxidation resistance \cite{steinbrueck2024overview}. This behaviour arises from the formation of a stable, protective $\mathrm{Al_2O_3}$ layer whose slow-growing, parabolic kinetics substantially delay runaway oxidation in accident scenarios \cite{steinbrueck2024overview}. Modern alloy development, particularly work led by Oak Ridge National Laboratory, has refined FeCrAl compositions to optimise corrosion resistance, manufacturability, and irradiation tolerance. Experimental and modelling studies show that FeCrAl claddings can extend reactor coping times and maintain structural integrity at temperatures well above the failure threshold of Zircaloy~\cite{Terrani2014,Zinkle2014}, reinforcing their potential for deployment in light-water reactors. The safety margin gains are explicitly quantified in~\cite{Terrani2014}, showing that much slower oxidation kinetics of iron-based alloys relative to Zircaloy directly extend the time available for emergency cooling intervention under beyond design basis accident scenarios (BDBA), while~\cite{Zinkle2014} provides an ATF systems perspective encompassing the full range of DBA and BDBA transients that the next generation cladding must be resistant to.

FeCrAl alloys are under consideration for deployment primarily in light-water reactors, both pressurised and boiling water designs, as a drop-in replacement for Zircaloy cladding tubes \cite{Field2018handbook}. The transition from laboratory demonstration to reactor deployment requires satisfying a demanding set of qualification criteria spanning corrosion performance in primary coolant, neutron economy penalties arising from the higher neutron absorption cross-section of Fe relative to Zr, weldability, pellet-cladding interaction behaviour, and irradiation performance over multi-year fuel cycles. Experimental evaluation of this irradiation performance has been ongoing, with tensile testing of model FeCrAl alloys neutron-irradiated to doses spanning all life stages of the cladding revealing that hardening increases with Cr content at doses below 2 dpa before saturating above 7 dpa due to $\alpha$-Cr phase \cite{Field2017}. Alloys with 18 wt \% Cr show brittle fracture at the highest doses \cite{Field2017}, demonstrating that composition-dependent defect physics influence macroscopic property changes. A key requirement for licensing is the ability to predict long-term material evolution through microstructural changes, phase stability, and creep under combined thermal and irradiation loading, and mechanistic models of defect transport of the kind developed here are a necessary input to such predictions, linking atomistic processes to the mesoscale models used in fuel performance codes.

Despite this promise, the defect physics of FeCrAl alloys under irradiation remain poorly understood. Neutron damage continuously produces vacancy-interstitial Frenkel pairs and the mobility and evolution of these defects govern swelling, creep, precipitation kinetics, and phase transformations central to the long-term performance of cladding materials. Vacancy production and transport underpin the principal forms of irradiation-induced degradation in reactor structural materials, with surviving vacancies coalescing into cavities to drive swelling and redistributing to biased sinks under stress to drive irradiation creep~\cite{Onimus2020}. Existing theoretical treatments have captured vacancy behaviour only in idealised or compositionally simplified environments, whereas real FeCrAl alloys contain complex local chemical fluctuations. These local environments can significantly alter vacancy binding and migration pathways, as well as collective behaviours such as clustering tendencies, trapping and percolation.

Density functional theory calculations have established that Al strongly influences vacancy formation and migration energetics in BCC Fe, with Al atoms at or near the migration saddle point raising barriers substantially relative to the pure Fe case \cite{Domain2002,Zhao2018prm}. Molecular dynamics simulations using EAM potentials have examined cascade damage morphology and short-range defect clustering, finding that Cr and Al both affect defect survival fractions and cluster size distributions following displacement events \cite{Field2015,Field2017}, mirroring the broader pattern observed in concentrated solid-solution alloys where increasing chemical disorder prolongs thermal spike lifetimes and reduces stable Frenkel pair survival \cite{Zhang2015natcomm}. Experimentally, neutron irradiation studies on model FeCrAl alloys have documented void swelling suppression and the formation of Cr-rich $\alpha'$ precipitates at and below 382$^\circ$C, with hardening response depending strongly on bulk Cr content where compositions above 8~wt\% Cr show $\alpha'$-dominated hardening while lower-Cr alloys are governed by dislocation loop accumulation \cite{Field2015}. Atom probe tomography and small-angle neutron scattering have further characterised the kinetics of $\alpha'$ precipitation as a function of composition and dose \cite{Briggs2017}. Vacancy migration is itself a mechanism for chemical mixing, since each hop swaps the vacancy with a neighbouring atom; sustained transport can therefore redistribute solute locally and assist precipitate nucleation, as shown for cation vacancies in disordered $FeCr_2O_4$~\cite{Hatton2023}. However, these studies have generally not been able to disentangle the individual contributions of constituting elements which migrate to the overall vacancy transport, motivating the species-resolved approach taken in this study.

Atomistic simulation methods provide access to quantities that are experimentally inaccessible: vacancy diffusivities in concentrated alloys, migration pathways through chemically heterogeneous environments, and the sensitivity of barriers to local composition that cannot be isolated experimentally. The reasoning chain from Density Functional Theory (DFT) through NEB barrier databases to kinetic Monte Carlo (KMC) diffusivities is established for pure iron by \cite{Fu2005}, whose \textit{ab-initio} treatment of vacancy and interstitial migration in $\alpha$-Fe provided the foundation on which subsequent alloy KMC frameworks build.

A broader theme motivating this work is compositional engineering, meaning deliberately adjusting alloy chemistry to shift defect transport into a more favourable regime. This principle is well established in the high-entropy alloy community, where extreme chemical disorder has been shown to suppress defect mobility through barrier roughening and localised energy landscape distortion, improving radiation tolerance relative to conventional alloys \cite{George2019,Xing2025,Xu2022acta, ElAtwani2019}. A complementary route to the same end immobilises irradiation-induced defects by tuning the degree of local lattice distortion, with the most severely distorted alloys exhibiting near-frozen defect motion and negligible irradiation-induced microstructural change \cite{Huang2026natcomm}. The mechanism has been traced to percolation effects: when the concentration of a faster-diffusing species approaches the lattice site percolation threshold, the connected network of low-barrier paths breaks down and the overall diffusivity drops below that of either pure component \cite{Osetsky2018,Osetsky2020}.

The present work applies a reduced-order kinetic Monte Carlo framework, parameterised by species-specific linear regression models trained on Nudged Elastic Band (NEB) calculations, to quantify vacancy diffusion in FeCrAl alloys across a systematic range of Fe/Cr compositions and temperatures from 500 K to 1500 K, relevant for normal operation and accident regimes. Emphasis is placed on the influence of Al on migration barriers despite its low concentration, and on identifying compositional regimes which strongly influence global vacancy transport rates, both of which have implications for FeCrAl alloy design and reactor lifetime modelling.

\section{Methodology}

All interatomic interactions are evaluated using the embedded atom method (EAM) potential for the Fe-Cr-Al ternary system developed by \cite{Liao2020}. The potential is fitted to a DFT database spanning the binary Fe-Cr, Fe-Al, and Cr-Al sub-systems as well as the ternary Fe-Cr-Al, and is validated against lattice parameters, elastic constants (bulk and shear moduli), self-interstitial atom (SIA) formation energies, vacancy formation and migration energies, stacking fault energies, and the binding energies of mixed and pure dumbbells (Fe-Cr, Fe-Al, Cr-Cr, Al-Al) in a base BCC Fe matrix \cite{Liao2020}.  This potential is selected over alternatives (e.g., MEAM \cite{LEE2001527} or earlier Finnis-Sinclair models \cite{Dai_2006}) because it is specifically refit to address shortcomings in defect properties, earlier potentials either misrepresented Fe-Cr mixing enthalpies at low Cr content or gave dumbbell binding energies in poor agreement with DFT \cite{Liao2020}.

\begin{figure*}[t]
    \centering
    \includegraphics[width=0.32\linewidth]{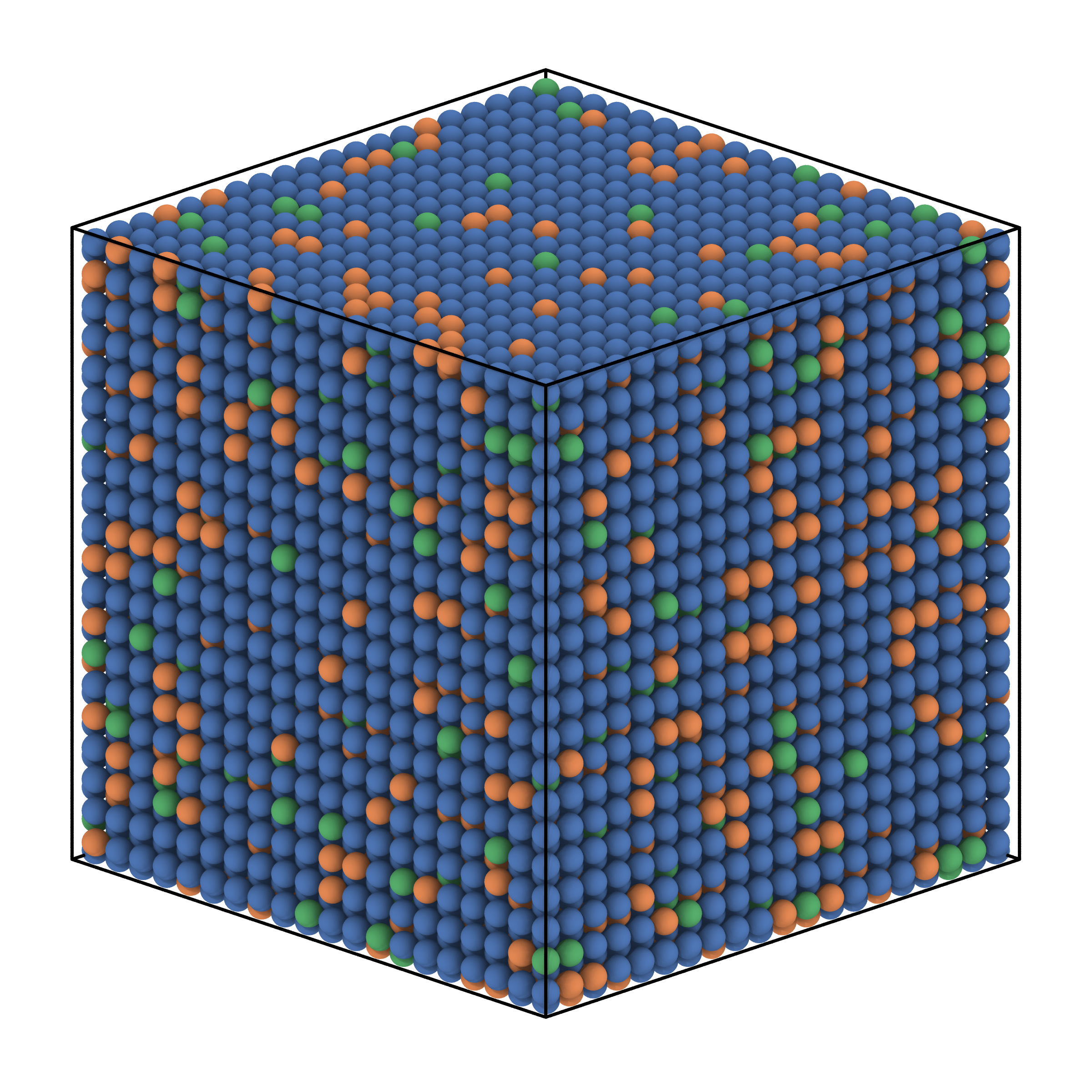}\hfill
    \includegraphics[width=0.32\linewidth]{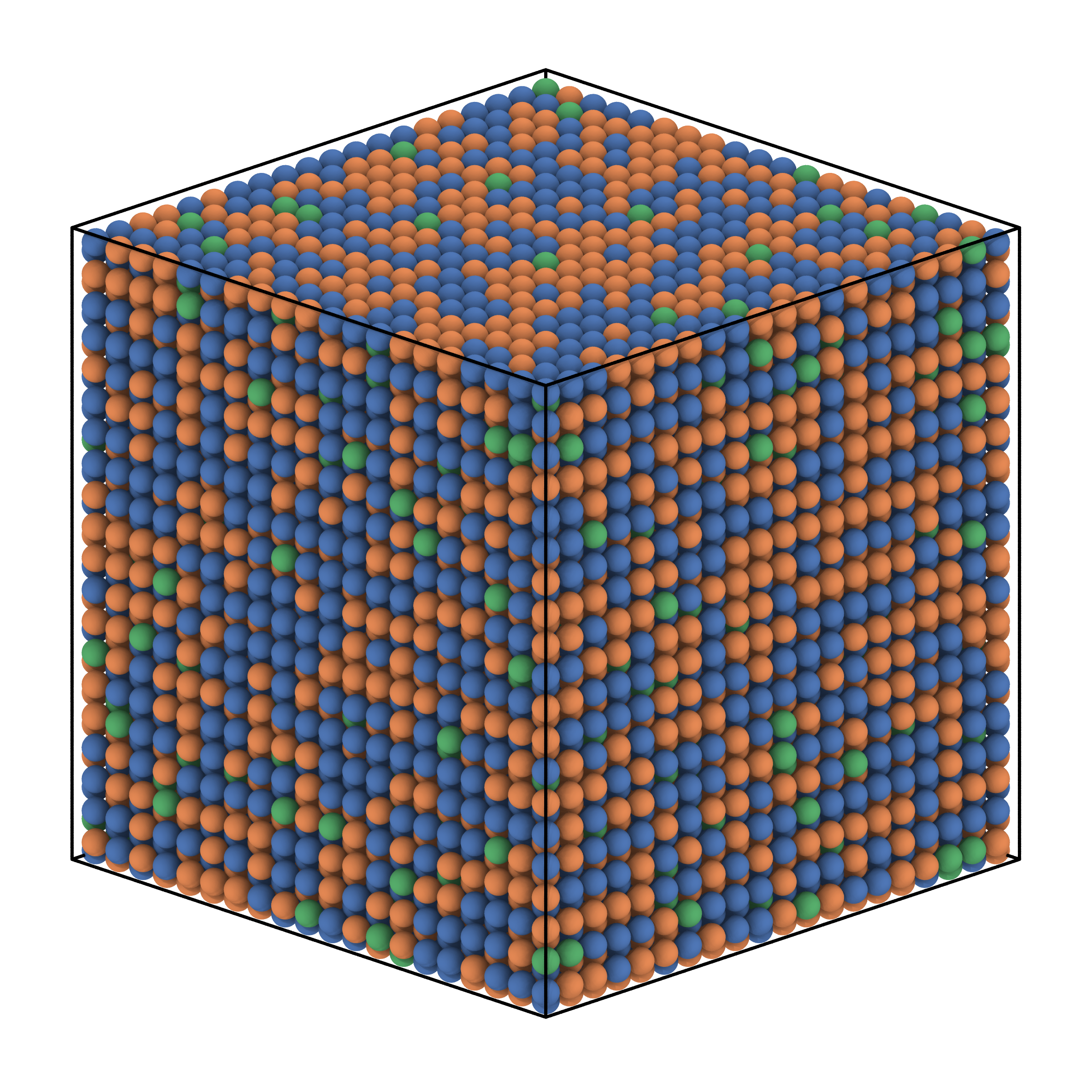}\hfill
    \includegraphics[width=0.32\linewidth]{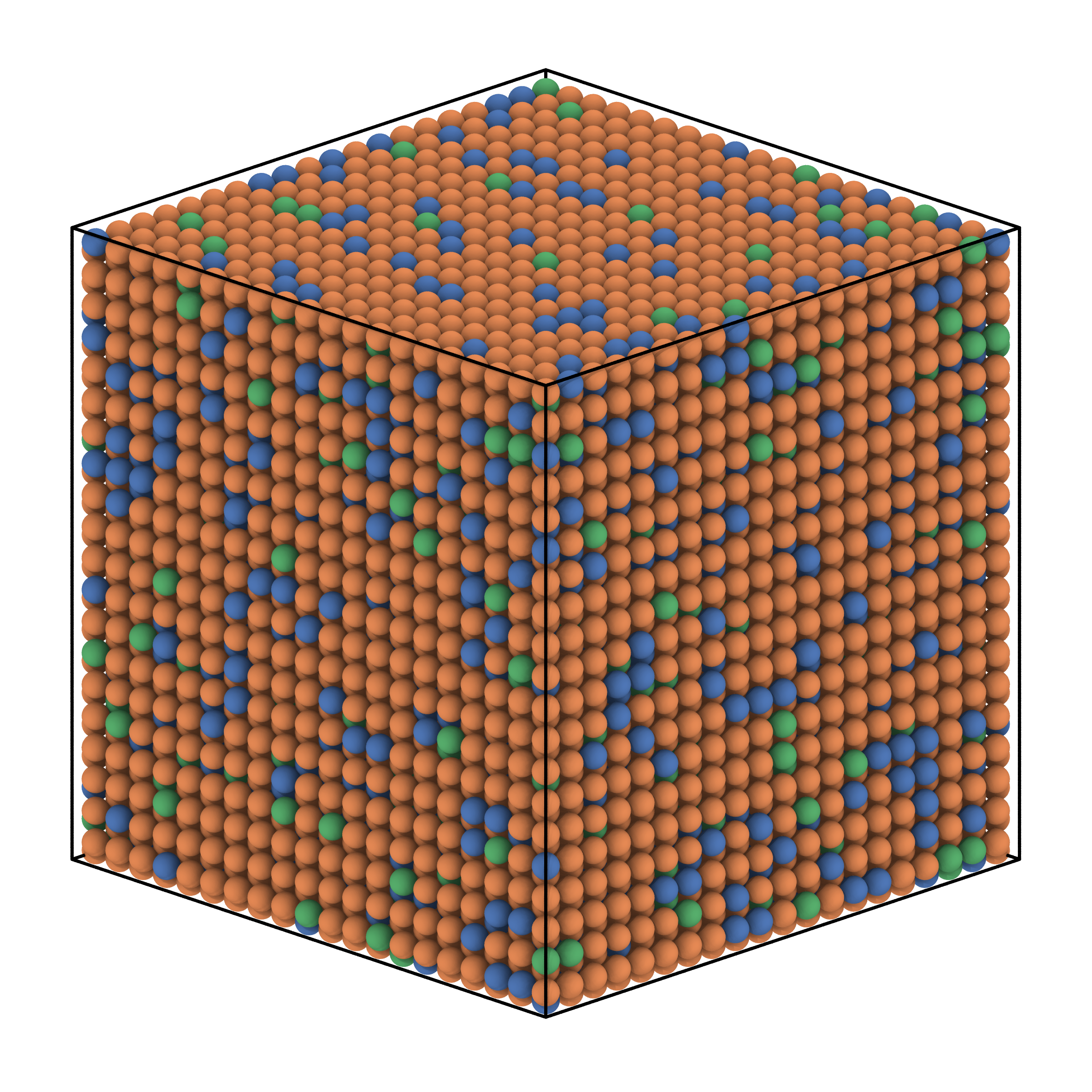}
    \caption{Representative decorated supercells at the Fe-rich (left), equiatomic Fe:Cr (centre), and Cr-rich (right) compositions, with Al fixed at $5\,at.\%$. Fe, Cr, and Al are shown in blue, orange, and green respectively. Each cell contains a single vacancy.}
    \label{fig:structure}
\end{figure*}

In all simulations, Al content is fixed at $5\,at.\%$, with the Fe:Cr ratio varied to produce five compositions spanning Fe$_{80}$Cr$_{15}$Al$_5$ to Fe$_{15}$Cr$_{80}$Al$_5$. Representative decorated cells are shown in Fig.~\ref{fig:structure}.

This Al concentration is representative of the compositions under development for ATF cladding applications, which typically contain $4-10\,at.\%$ Al \cite{Field2018handbook,Yamamoto2015}. It also preserves the single-phase BCC solid solution that is the focus of this work due to preliminary simulations at higher Al contents caused the lattice to collapse towards an FCC-like arrangement, consistent with the tendency of Al to stabilise close-packed structures at elevated concentrations. Since the present work is concerned with vacancy transport in the BCC phase relevant to normal and accident operating conditions, compositions are restricted to the regime in which this structural stability is guaranteed.

Vacancy migration data is generated using the \textit{Hop-Decorate} workflow \cite{Hatton2025HopDec} in which a BCC Fe reference structure is initialised and Fe, Cr, and Al atoms are randomly assigned to lattice sites according to a target composition and the minimum energy paths between initial and final vacancy configurations are computed using the climbing-image nudged elastic band (CI-NEB) method \cite{Henkelman2000Tangent, Henkelman2000CINEB}. This data generation is conducted for global compositions of Fe$_{80}$Cr$_{15}$Al$_5$, Fe$_{47.5}$Cr$_{47.5}$Al$_5$, and Fe$_{15}$Cr$_{80}$Al$_5$. Hops of all three species (Fe, Cr, Al) are sampled across a range of local chemical environments by repeating this procedure for many distinct vacancy sites and supercell decorations, yielding a training database of energy barriers that broadly samples the compositional disorder of the alloy. To obtain statistical sampling of the disordered local environment without altering the global composition, each composition is independently re-decorated 1500 times by varying the random seed, where decoration refers to the random assignment of species to lattice sites.

The local chemical environment of each hop is encoded using a descriptor that counts the number of Fe, Cr, and Al atoms within fixed radial shells around two geometrical reference points: the saddle-point between the initial and final vacancy positions, and the migrating atom in both its initial and final positions. The saddle-point counts characterise the environment through which the atom passes, while the difference between initial and final counts around the migrating atom captures the chemical asymmetry of the hop. This difference is interpreted as whether the atom is moving towards or away from regions enriched in a given species. An exploratory descriptor is first constructed using four NN shells around the saddle-point minima and five NN shells around the migration paths, yielding 27 features per hop, in order to assess the relative importance of each shell. Separate linear surrogate models are fitted by ordinary least-squares regression for each migrating species (Fe, Cr, Al), since a single combined model produced substantially higher errors. For each species, the fitted coefficients are ranked by absolute magnitude, and reduced models retaining progressively fewer features are refitted to identify the descriptors contributing most to the barrier. This analysis motivates truncation to three NN shells around each reference point, yielding 18 descriptors per hop. Since all features are integer atom counts the coefficient magnitude is a way to gauge the contribution of a given term.

The choice of an unregularised linear model with manual feature selection is a deliberate trade-off. Non-linear neural-network surrogates can in principle achieve lower prediction error in complex alloys \cite{Huang2025,Xu2022acta, Talapatra2025}, but require additional parameters, larger training datasets, and sacrifice the physical interpretability of the fitted coefficients. With only a handful of descriptors retained per species, magnitude-based selection is both tractable and transparent, while keeping each barrier evaluation a single dot product that is cheap to evaluate at every kMC step. The broader strategy of replacing NEB calculations with an environment-based regression follows the precedent set by Castin \textit{et al.} \cite{Castin2014}, who showed that surrogate models trained on a NEB database can reproduce bulk diffusivities. The present work adopts the same philosophy with a linear model in place of a neural network.

KMC simulations are performed using a residence-time algorithm \cite{Voter2007} widely used in radiation damage work \cite{Fu2005}. The residence-time algorithm selects events with the probability proportional to the exponential of their rate and advances the simulation clock. At the start of each simulation, a single vacancy is introduced at a fixed lattice site and its first-nearest neighbours identified. At each KMC step, the surrogate model is evaluated for each vacancy-neighbour pair to obtain the hop barrier, and the corresponding rate computed as:

\begin{equation}
  \Gamma = \nu_0 \exp\!\left(-\frac{E_b}{k_B T}\right),
\end{equation}

with attempt frequency $\nu_0 = 10^{13}$\,s$^{-1}$. This rate expression follows directly from the harmonic transition state theory approximation \cite{Henkelman2000CINEB}, which is well justified for tightly-packed crystalline solids at temperatures well below their melting point. Each trajectory was run for $10^5$ KMC steps, which is found to be sufficient to sample the diffusive regime across the temperature range studied. Simulations are performed at temperatures of 500, 700, 900, 1100, 1300, and 1500 K, spanning the range from normal LWR operating conditions through to temperatures relevant to loss of coolant accident (LOCA) scenarios. 

Vacancy coordinates are unwrapped across periodic boundaries at each step to obtain a continuous trajectory, and the squared displacement from the initial position computed at each KMC step. The mean squared displacement is averaged across the five independent trajectories per composition-temperature combination, and the diffusion coefficient is extracted from its gradient. The standard deviation across trajectories is taken as the statistical uncertainty on $D$. Activation energies $E_a$ and pre-exponential factors $D_0$ are then obtained by fitting the temperature-dependent diffusivities to the Arrhenius equation:

\begin{equation}
  D = D_0 \exp\!\left(-\frac{E_a}{k_B T}\right).
\end{equation}

\section{Results \& Discussion}

\subsection{Migration Energies in Pure End-members}

\begin{figure}[!tb]
    \centering
    \includegraphics[width=\linewidth]{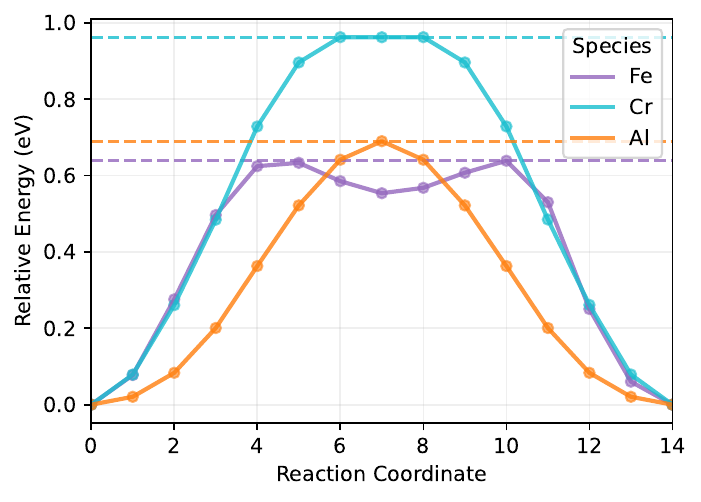}
    \caption{Migration energy of a vacancy defect in the pure, bulk materials. Dashed lines indicate the maximum of energy barriers.}
    \label{fig:end}
\end{figure}

Figure~\ref{fig:end} shows the migration energy profiles for a vacancy hop in the three pure end-member environments. The barrier heights differ substantially across species: Vacancies in pure Fe have a migration barrier of approximately $0.63$ eV, Cr approximately $0.95$ eV, and Al approximately $0.69$ eV, establishing the ordering $E_b^\text{Fe} < E_b^\text{Al} < E_b^\text{Cr}$ that recurs throughout the analysis. We note that pure Al is calculated in the FCC structure, as this is its stable phase at ambient conditions, whereas Fe and Cr are calculated in BCC. The profile shapes also differ qualitatively. The Cr and Al profiles have a single saddle-point, whereas the Fe profile exhibits a shallow local minimum representing a so-called split-vacancy. This dip is small in magnitude and therefore unlikely to trap the vacancy for a dynamically significant duration, rather it is consistent with a slight asymmetry in the Fe-Fe interaction across the migration path. These pure end-member barriers serve as a reference point for the alloy results that follow.

\subsection{Data Generation}

\begin{figure}[!tb]
    \centering
    \includegraphics[width=\linewidth]{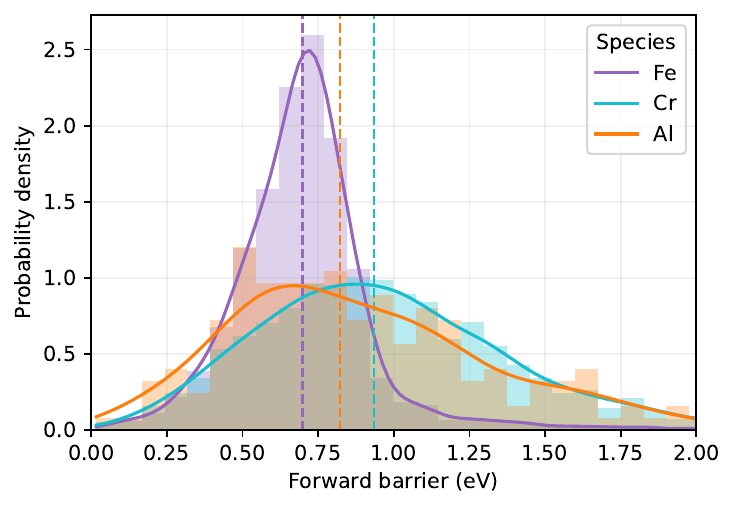}
    \caption{Distribution of energy barriers by migrating species with dashed lines indicating the median value.}
    \label{fig:hist}
\end{figure}

Figure~\ref{fig:hist} shows the combined distributions of energy barriers grouped by migrating species across the sampled global compositions (Fe$_{80}$Cr$_{15}$Al$_5$, Fe$_{47.5}$Cr$_{47.5}$Al$_5$, and Fe$_{15}$Cr$_{80}$Al$_5$) from the Hop-Decorate workflow. All three distributions are plotted as probability densities rather than raw counts so that the relative shapes can be compared on equal footing. Interestingly, the ordering of the median values is the same as the pure end-members shown in Figure~\ref{fig:end}, though the median values are lower than the pure cases. Additionally, by inspecting the reaction pathways we find that the shallow intermediate for Fe migration is not present when the local environment contains chemical complexity indicating that the local environment not only modulates barrier height but can change details of the migration pathway.

It can be seen that Fe distribution is relatively narrow and centred around $\simeq$ 0.75~eV, indicating that Fe migration barriers are comparatively insensitive to the precise local chemical arrangement. The Cr and Al distributions are substantially broader, spanning from  $\simeq$ 0.25~eV to $>$ 1.5~eV, demonstrating that local environment significantly affects migration of these species. The breadth of the Cr distribution in particular reflects the strong dependence of Cr migration on whether the hop is directed toward or away from specific species, as discussed further in the surrogate model analysis below.

The Al distribution is different due to the training set for Al containing far fewer events than for Fe or Cr, a direct consequence of the $5~at.\%$ Al constraint. At this concentration, first-neighbour hops involving an Al migrator are statistically rare relative to Fe and Cr events. The limited Al statistics are an inherent feature of the
chosen composition regime rather than a sampling deficiency, and the resulting larger uncertainty in the Al surrogate model is carried through to the discussion.

\subsection{Surrogate Model Fitting}

Separate linear surrogate models are fitted for each migrating species using the barrier training database. We find that fitting a single combined model for all three species produces higher errors confirming that the identity of the migrating species is itself a primary determinant of the barrier and that separate species-resolved expressions are physically necessary. The resulting barrier expressions are described in equations (3), (4), and (5).

The equations contain two physically distinct classes of terms. The $N_\beta^{r^S_k}$ terms count atoms of species $\beta$ within the $k$th radial (r) shell of the saddle-point, which is the geometric midpoint between the vacancy and the migrating atom. The $\Delta N_\beta^{r^M_k}$ terms capture the asymmetry of the composition on either side of the hop defined as $\Delta N_\beta = N_\beta^\text{final}- N_\beta^\text{initial}$. A positive $\Delta N_{\beta}$ means the vacancy is moving into an environment enriched in $\beta$. A negative coefficient on such a term therefore means that moving towards that species is energetically favourable. The final term is a correction and is related to the energy barrier of the vacancy in the corresponding pure materials. Indeed, the order in the magnitude in these corrections is the same as the in Figure \ref{fig:end}.

\begin{widetext}
\noindent
\begin{minipage}{0.48\textwidth}
\begin{equation}\label{eqn-barrier-Fe}
E_b^{\text{Fe}} = \begin{array}{l}+ 0.2879\, N_{\text{Fe}}^{r^S_1} + 0.1865\, N_{\text{Cr}}^{r^S_1} + 0.3363\, N_{\text{Al}}^{r^S_1} \\
+ 0.0519\, N_{\text{Fe}}^{r^S_3} + 0.0760\, N_{\text{Cr}}^{r^S_3} + 0.0423\, N_{\text{Al}}^{r^S_3} \\
+ 0.0655\, \Delta N_{\text{Cr}}^{r^M_1} - 0.1617\, \Delta N_{\text{Fe}}^{r^M_3} \\
- 0.1804\, \Delta N_{\text{Cr}}^{r^M_3} - 0.1771\, \Delta N_{\text{Al}}^{r^M_3} - 2.3951
\end{array}
\end{equation}
\end{minipage}
\hfill
\begin{minipage}{0.48\textwidth}
\begin{equation}\label{eqn-barrier-Cr}
E_b^{\text{Cr}} = \begin{array}{l}
+ 0.3228\, N_{\text{Fe}}^{r^S_1} + 0.2809\, N_{\text{Cr}}^{r^S_1} + 0.3962\, N_{\text{Al}}^{r^S_1} \\
+ 0.0751\, N_{\text{Al}}^{r^S_2} + 0.0277\, N_{\text{Cr}}^{r^S_3} \\
- 0.0446\, \Delta N_{\text{Fe}}^{r^M_1} - 0.2749\, \Delta N_{\text{Al}}^{r^M_2} \\
- 0.2906\, \Delta N_{\text{Fe}}^{r^M_3} - 0.3054\, \Delta N_{\text{Cr}}^{r^M_3} \\
- 0.1785\, \Delta N_{\text{Al}}^{r^M_3} - 1.5212
\end{array}
\end{equation}
\end{minipage}

\vspace{1em}

\begin{center}
\begin{minipage}{0.48\textwidth}
\begin{equation}\label{eqn-barrier-Al}
E_b^{\text{Al}} = \begin{array}{l}
+ 0.4321\, N_{\text{Fe}}^{r^S_1} + 0.3272\, N_{\text{Cr}}^{r^S_1} + 0.5827\, N_{\text{Al}}^{r^S_1} \\
- 0.2458\, N_{\text{Fe}}^{r^S_2} - 0.1269\, N_{\text{Al}}^{r^S_2} + 0.1184\, N_{\text{Fe}}^{r^S_3} \\
+ 0.0777\, \Delta N_{\text{Fe}}^{r^M_1} + 0.3524\, \Delta N_{\text{Fe}}^{r^M_3} \\
+ 0.3952\, \Delta N_{\text{Cr}}^{r^M_3} + 0.2745\, \Delta N_{\text{Al}}^{r^M_3} - 1.6896
\end{array}
\end{equation}
\end{minipage}
\end{center}
\end{widetext}

A consistent feature in these equations is that Al carries the largest first nearest neighbor (1NN) saddle-point coefficient in all three models (0.3363 for Fe migrating, 0.3962 for Cr, and 0.5827 for Al). Each Al atom present at the 1NN saddle shell raises the Fe migration barrier by $+0.048$~eV, the Cr barrier by $+0.073$~eV, and the Al barrier by $+0.151$~eV. This disproportionate influence of Al at the saddle is consistent with the broader observation that Al strongly impedes vacancy migration in BCC systems even at dilute concentrations and is supported by DFT calculations in studies of minor-element doping effects in concentrated alloys, where Al introduction is shown to raise vacancy migration energies relative to the undoped matrix \cite{Zhao2022jnm}. More broadly, the sensitivity of migration barriers to local chemical coordination, captured here through the saddle-point shell counts, is a defining feature of the defect energy landscape in all concentrated solid-solution alloys, as extensively reviewed in \cite{Zhang2022chemrev}.

Among the migration-shell terms $\Delta N^{r^M_n}$, the 3NN coefficients consistently carry a large absolute weight across all three migrators. This is consistent with prior atomistic studies of vacancy migration in BCC Fe-based alloys, which show that the influence of the local environment extends well beyond the immediate first-neighbour shell. DFT-NEB calculations on Fe-Cr demonstrate that the vacancy migration barrier continues to evolve as second-nearest-neighbour sites are progressively occupied with Cr, leading to the conclusion that any predictive model must encode not just the change in configurational energy but also an explicit description of the saddle-point chemistry and its surroundings~\cite{COSTA2014425}. A complementary study deriving exact transport coefficients for dilute BCC Fe-$X$ alloys finds it necessary to extend the kinetic interaction shell out to 5NN, on the grounds that migration paths traversing 2NN-4NN-3NN-4NN sequences contribute non-negligibly to vacancy-solute coupling and that models truncated at 1NN or 2NN systematically miss part of the kinetic response~\cite{PhysRevB.90.104203}.

Furthermore, the 3NN shell lies outside the bond-breaking region directly around the migrating atom and the vacancy but is still close enough that differing species at this range change the local environment along the migration path. The inner shells are already accounted for by separate features and the outer shells are too distant to contribute strongly, so the 3NN terms are left to carry the part of the barrier variation that depends on the wider environment. That this wider environment matters at all is a physical rather than a geometric question, addressed in the literature through the combined effects of chemical interactions, magnetism and elastic perturbations beyond the first neighbour shell~\cite{COSTA2014425, PhysRevB.90.104203}. Pulling apart the individual contributions from chemistry, elasticity and magnetism within each shell falls outside what a linear surrogate model can do and would require targeted DFT or atomistic sensitivity calculations of the kind reported in those works.

Additionally, the directional preferences of Fe and Al migrators are directly opposed with respect to Cr. That is, hopping toward a Cr-enriched environment lowers the Fe migration barrier by $-0.180$~eV but raises the Al barrier by $+0.395$~eV. This means that under the same local configuration, the Fe vacancy is \lq attracted' toward Cr-rich regions while the Al vacancy is \lq repelled' from them, implying that Fe and Al defects preferentially explore different compositional environments during a KMC trajectory reflecting a kinetic preference for species segregation, which is consistent with the formation of a Cr-rich phase/species segregation \cite{Field2018handbook}. An additional feature specific to the Cr model is a strong sensitivity to Al asymmetry at 2NN: $\Delta N_\text{Al}^{r^M_2} = +1$ lowers the Cr barrier by $0.275$~eV, the largest single 2NN effect in any of the three expressions and one with no counterpart in the Fe or Al models.

\subsection{Vacancy Diffusivity}

\begin{figure}
    \centering
    \includegraphics[width=\linewidth]{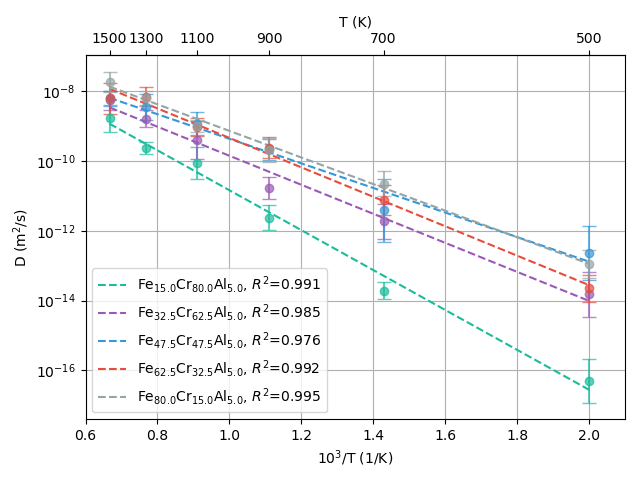}
    \caption{Arrhenius plot of the vacancy diffusion coefficient $D$ as a function of inverse temperature for the five compositions studied.}
    \label{fig:Diffusivity}
\end{figure}

The previously defined surrogate models are implemented in a surrogate-driven KMC model to derive the long range diffusivities associated within each global composition. Figure~\ref{fig:Diffusivity} shows the diffusivity of vacancies in all five alloy compositions. All compositions exhibit linearity in $\ln(D)$ vs $1/T$ across the full temperature range of 500-1500~K, with $R^2 > 0.97$ in all cases, confirming that vacancy diffusion follows thermally activated Arrhenius behaviour throughout. The error bars in Figure~\ref{fig:Diffusivity} originate from the redecoration. Each data point is averaged over multiple statistically independent initial random assignments of Fe, Cr and Al atoms to the cell at the same nominal composition. The bars therefore measure how much the extracted diffusivity changes when the global atomic arrangement is changed, while overall composition and temperature are held fixed. The magnitude of the error increases as temperature decreases which is consistent with a strongly heterogeneous energy landscape where local trapping is most dominant at lower temperature.

\begin{table}
\centering
\label{tab:arrhenius}
\begin{tabular}{lcc}
\hline
Composition & $E_a$ (eV) & $D_0$ (m$^2$\,s$^{-1}$) \\
\hline
Fe$_{80}$Cr$_{15}$Al$_5$   & 0.756 & $4.76 \times 10^{-6}$ \\
Fe$_{62.5}$Cr$_{32.5}$Al$_5$ & 0.838 & $7.70 \times 10^{-6}$ \\
Fe$_{47.5}$Cr$_{47.5}$Al$_5$ & 0.699 & $1.45 \times 10^{-6}$ \\
Fe$_{32.5}$Cr$_{62.5}$Al$_5$ & 0.826 & $2.07 \times 10^{-6}$ \\
Fe$_{15}$Cr$_{80}$Al$_5$   & 1.135 & $7.70 \times 10^{-6}$ \\
\hline
\end{tabular}
\caption{Arrhenius parameters extracted from KMC simulations for each 
alloy composition.}
\end{table}

We clearly see a strong dependence of D on global composition. This can be seen most clearly in the extracted activation energies and pre-exponential factors summarised in Table I. The activation energies are non-monotonic from Fe-rich to Cr-rich converging to $\simeq0.70-0.84~eV$, implying that the threshold global concentration for Fe percolation is $15\% - 32.5\%$ and beyond this threshold, Fe is sufficiently connected in the lattice that vacancy migration is dominated by Fe migration. These findings are consistent with Fe having the lowest energy barrier (fastest migration rate) in Figures \ref{fig:end} and \ref{fig:hist}, a trend captured by the surrogate models. While a trend is hard to distinguish due to relatively large error bars, there is a roughly monotonic increase in overall diffusivity as global Fe concentration increases which may reflect the decrease in sluggish migration events involving Cr.

Below this threshold the Cr-rich composition (Fe$_{15}$Cr$_{80}$Al$_5$) shows a markedly elevated activation energy of 1.135~eV and an overall diffusivity that falls several orders of magnitude below the Fe-rich cases at lower temperatures. For example, at the LWR operating temperature of $\sim$570~K, this composition exhibits diffusivities as low as $\sim10^{-17}$~m$^2$\,s$^{-1}$, compared to $\sim10^{-13}$~m$^2$\,s$^{-1}$ for Fe-rich alloys. This is consistent with a regime in which vacancy migration is dominated by Cr hops, leading to so-called sluggish diffusion. This slow diffusion rate in Cr-rich compositions could have significant impact on defect annihilation rates and vacancy clustering in Cr-rich $\alpha'$ phases known to form under irradiation of Fe-rich FeCrAl alloys \cite{Field2018handbook}.

\begin{figure}[t]
    \centering
    \includegraphics[width=\linewidth]{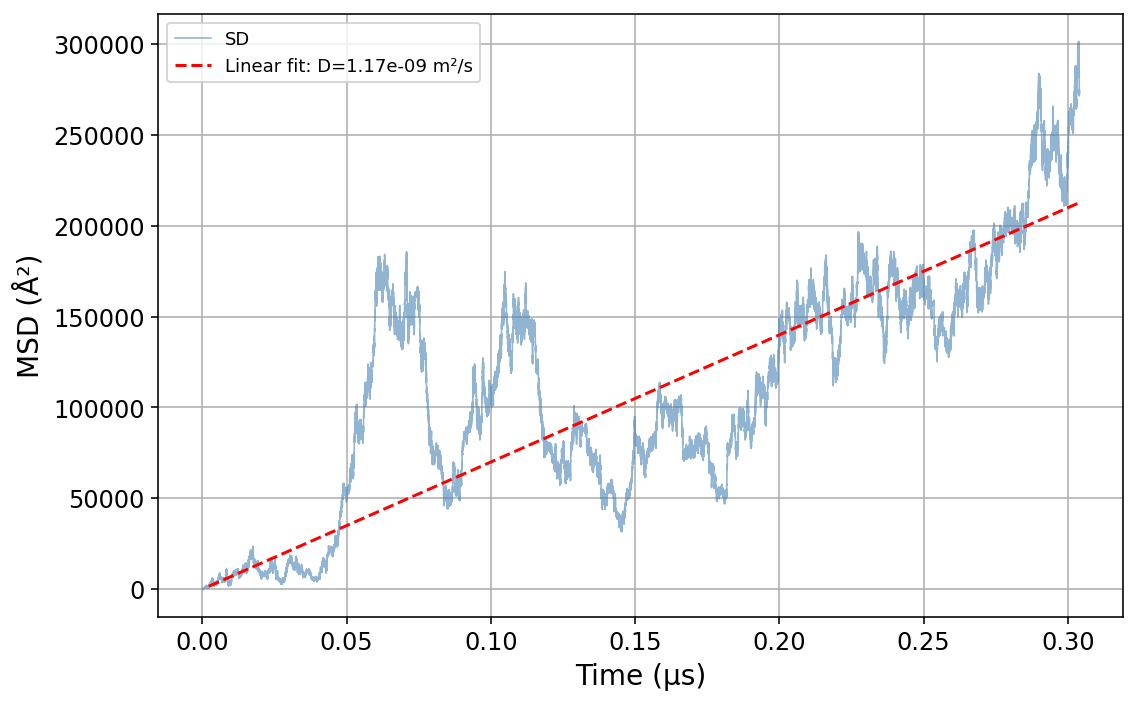}
    \caption{Representative mean squared displacement as a function of
    time for the Fe$_{80}$Cr$_{15}$Al$_5$ composition at 1100~K.
    Individual trajectories are shown in grey; the ensemble mean is
    shown in colour. The dashed line shows the linear fit to the
    diffusive regime used to extract $D$.}
    \label{fig:msd}
\end{figure}

Figure~\ref{fig:msd} shows a representative mean squared displacement (MSD) trajectory derived from a single KMC run. Clearly these trajectories are not linear. The squared displacement climbs to nearly $2 \times 10^{5}$~\AA$^2$ within the first $0.7 \times 10^{-7}$~s, then drops back below $5 \times 10^4$~\AA$^2$ shortly after, surges again near $1.1 \times 10^{-7}$~s, collapses once more around $1.5 \times 10^{-7}$~s, and only in the second half of the simulation does it begin to adopt a roughly linear regime. 

A random walk in a homogeneous lattice would produce an increasing MSD with a slope corresponding to the single hop. However, each peak in the plot represents the vacancy escaping the current neighbourhood and executing a sequence of hops along a network of low-barrier sites. Each downturn then indicates the percolation network terminating, where the vacancy re-enters a region of higher barriers, in which it becomes confined and executes repeated forward and backward hops. The fact that the squared displacement can decrease significantly on these timescales is a consequence of the trajectory of such defects through a chemically rough potential energy landscape. This trajectory consists of a small number of dominant escape routes through the cell, separated by trapping pockets in which the vacancy oscillates without net progress.

It should be noted, however, that for each global composition and each temperature, a total of 10 trajectories of the type shown in Figure \ref{fig:msd} are collected, averaged and then a linear fit produced. The diffusion coefficient is then extracted from the gradient of this linear fit via $D = \langle r^2 \rangle / 6t$.

\section{Discussion}

The primary motivation for developing accident tolerant fuel cladding materials is their ability to maintain structural integrity under both normal operating conditions and the extreme thermal excursions associated with (LOCA) scenarios. The results presented here allow the vacancy diffusion behaviour of FeCrAl alloys to be examined across both regimes.

Under normal light water reactor operating conditions, cladding temperatures are typically in the range of $550-600~K$. Inspection of Figure~\ref{fig:Diffusivity} reveals that at these temperatures, four of the five compositions exhibit vacancy diffusivities that are indistinguishable within statistical uncertainty, spanning a narrow range of $\sim10^{-13}$--$10^{-12}$~m$^2$\,s$^{-1}$. This insensitivity is consistent with observations in binary NiFe alloys, where the sluggish diffusion effect peaks near the percolation threshold and modest compositional deviations produce relatively small changes in diffusivity \cite{Osetsky2021}, and with experimental evidence that compositional sensitivity to barrier fluctuations is most pronounced at low temperatures \cite{Tsai2013,Kottke2020}. In FeCrAl, Fe-dominated transport across the Fe-rich and equiatomic compositions provides an analogously flat diffusivity response until the Cr-rich region is approached. The Cr-rich composition Fe$_{15}$Cr$_{80}$Al$_5$ deviates significantly from this trend, exhibiting diffusivities approximately two orders of magnitude lower.

The temperature conditions change substantially under accident conditions. During a LOCA event, cladding temperatures can rise rapidly to $1200$ K and above, a regime in which the compositional dependence of vacancy diffusivity becomes pronounced. The Cr-rich alloy maintains its suppressed diffusivity relative to the Fe-rich compositions, with an activation energy of $1.135~eV$ compared to $0.70-0.84~eV$ for the remaining alloys. In the context of radiation damage, vacancy diffusivity governs the rate at which vacancy-interstitial recombination occurs. The faster-moving vacancies are more likely to encounter and annihilate with interstitials before either defect reaches a sink. Under accident-relevant temperatures where thermal diffusion dominates, a higher activation energy suppresses vacancy mobility, reducing the rate at which vacancies migrate to grain boundaries and dislocation sinks and thereby raising the local vacancy-interstitial encounter probability. This mechanism is consistent with the percolation framework of \cite{Xu2022matdes}, in which the potential energy landscape is the dominant factor governing defect transport at elevated temperature. Fe$_{15}$Cr$_{80}$Al$_5$ therefore presents a compelling case for further investigation as an accident tolerant cladding material, combining the known oxidation resistance of Cr-rich FeCrAl with favourable vacancy kinetics.

It is worth noting that a vacancy-only treatment does not address interstitial cluster transport. Experimental work on concentrated Ni-based alloys shows that compositional complexity suppresses the long-range glide of interstitial clusters and converts their motion to a more three-dimensional character, increasing vacancy-interstitial recombination rates \cite{Lu2016natcomm}. A further composition-dependent process not captured here is the radiation-induced transformation of $\tfrac{1}{2}\langle 111\rangle$ to $\langle 100\rangle$ dislocation loops in Fe-Cr alloys, which is strongly sensitive to Cr concentration \cite{Zhang2021jnm}. At the high Cr content of Fe$_{15}$Cr$_{80}$Al$_5$, this loop-type transition may affect the effective defect sink density and hence the vacancy recombination rate independently of the migration kinetics studied here.

A further limitation is the restriction of the local environment descriptor to atom counts within fixed radial shells: while this linear framework captures the dominant trends in barrier variation, it cannot encode directional or symmetry-breaking effects in the local environment, nor account for magnetic contributions to the energy landscape which are known to be significant in Fe-rich BCC alloys. A related limitation concerns jump correlation effects, since a faithful description of diffusion in multi-principal element alloys requires the full energy landscape \cite{Thomas2020}, and the present model does not produce the correlated back-jumping that occurs when a vacancy crosses a low barrier. The use of a classical EAM potential for NEB calculations, while computationally necessary at this scale, also introduces systematic errors relative to DFT that have not been fully quantified for the migration barriers specifically. Validation of the surrogate model predictions against DFT-computed barriers for a representative subset of configurations would strengthen confidence in the quantitative results.

Finally, the KMC simulations treat the alloy as a static, frozen lattice with no account taken of local relaxation around the vacancy or compositional evolution during irradiation. In a real irradiation environment, radiation-induced segregation would progressively alter the local composition around defect sinks, modifying the effective barrier landscape over time, and would generate short-range chemical order around defect clusters that further roughens the local energy landscape and modifies defect migration rates beyond what a static random-alloy model captures \cite{Cao2021}. Extending the present framework to incorporate composition evolution under irradiation represents a natural direction for future work. An extension would need to account for Cr and Al segregation to grain boundaries and dislocation loops documented experimentally in irradiated FeCrAl \cite{Field2015}, as well as the multi-year timescales over which composition-dependent microstructural evolution governs cladding performance in commercial LWR fuel cycles \cite{Pint2015,Zinkle2014}.

\section{Conclusion}

This work presents a species-resolved kinetic Monte Carlo framework for vacancy diffusion in FeCrAl alloys, parameterised by linear surrogate models trained on a database of CI-NEB migration barriers generated through the HopDec workflow. Five compositions spanning Fe$_{80}$Cr$_{15}$Al$_5$ to Fe$_{15}$Cr$_{80}$Al$_5$ are examined across temperatures from $500~K$ to $1500~K$, covering both normal LWR operating conditions and elevated temperatures characteristic of LOCA scenarios.

The physical insight is that vacancy mobility in FeCrAl is governed by the local chemistry of individual hops rather than by global composition alone. Al exerts an influence on migration barriers that is disproportionate to its dilute concentration, and the distinct directional preferences of the three migrating species mean that the alloy cannot be treated as a single medium. This species-resolved picture explains why the macroscopic diffusivity is not recovered by a composition-weighted average of the pure end-member barriers, and reinforces the broader principle that defect transport in concentrated alloys is set by the heterogeneity of the local energy landscape.

Across most of the alloy range studied, vacancy diffusivity is comparatively insensitive to composition, with the Cr-rich regime standing apart as Fe$_{15}$Cr$_{80}$Al$_5$ exhibits a higher activation energy compared to the other compositions and diffusivities several orders of magnitude lower at the bottom of the temperature range. This separation, which accounting for error bars is lower in the normal operating temperature range becomes more pronounced under accident conditions, indicating that the Fe:Cr ratio becomes a meaningful design layer in the operating regime where the cladding structure matters most.

More broadly, the present work demonstrates that compositional engineering of FeCrAl can be guided by atomistic transport modelling rather than empirical trial. The approach generalises to other concentrated alloys wherever a migration energy barrier dataset can be generated and provides a foundation for the coupled composition and evolution problems which govern long-term cladding performance under irradiation. In identifying a Cr-rich composition that combines established oxidation resistance with vacancy kinetics, the present work points towards a testable direction for the next generation of accident-tolerant cladding alloys.

\begin{acknowledgements}
The authors acknowledge Fei Gao for providing the interatomic potential used in this work.
The authors thank Andrew Davies for providing a critical review of this manuscript.
The authors also acknowledge the NTEC master's programme for initiating this collaboration.
Finally, the authors jointly acknowledge the Cirrus HPC owned and operated by EPCC and the Hades computing cluster at Amentum for all computational work conducted in this study as well as Jim Skelton, Liam Marsden \& Andrew Dean for their help with this resource.
\end{acknowledgements}

\nocite{*}
\bibliography{myrefs}

\end{document}